\documentclass[aps,pre,preprint,showpacs]{revtex4-1}

\usepackage{epsfig}
\usepackage{graphics}
\usepackage{graphicx}
\usepackage{epstopdf}
\usepackage[dvipsnames]{xcolor}
\DeclareGraphicsExtensions{.jpeg,.eps,.pdf}

\begin{document}

\title{Phase diagram of quantum fluids. The role of the chemical potential and the phenomenon of condensation.}

\author{V\'{\i}ctor Romero-Roch\'{\i}n} \email{romero@fisica.unam.mx}

\affiliation{Instituto de F\'{\i}sica. Universidad Nacional Aut\'onoma de M\'exico. \\
Apartado Postal 20-364, 01000 M\'exico, D.F., Mexico}

\date{\today}

\begin{abstract}
We discuss the generic phase diagrams of pure systems that remain fluid near zero temperature. We call this phase a {\it quantum} fluid. We argue that the signature of the transition is the change of sign of the chemical potential, being negative in the normal phase and becoming positive in the quantum fluid phase. We show that this change is characterized by a phenomenon that we call {\it condensation}, in which a macroscopic number of particles is in their {\it own many-body ground state}, a situation common to Fermi and Bose gases. We show that the ideal Bose-Einstein Condensation fits in this scenario, but that it also permits the occurrence of a situation that we may call ``Fermi-Dirac Condensation". In addition, we argue that this phenomenon is also behind the development of superfluid phases in real interacting fluids.  However, only interacting systems may show the change from a thermal fluid to a quantum one as a true phase transition. As a corollary, this argument shows the necessity of the appearance of a ``supersolid" phase. We also indicate how these ideas may be useful in the description of of experiments with ultracold alkali gases.
\end{abstract}


\maketitle

\section{Introduction}

The resolution of the thermodynamics of interacting systems at very low temperatures starting from first principles is an aged endeavour that has even attracted the attention of great minds in the last and present Centuries, sparked early by the striking properties of Helium\cite{Kapitza,Allen,Osheroff,Balibard} and renewed with the spectacular experiments involving ultracold Alkaline gases\cite{Cornell,Ketterle,Hulet}. Initiating with the crucial experimental and theoretical works by Kapitza\cite{Kapitza}, Allen and Wisener\cite{Allen}, London\cite{London}, Tisza\cite{Tisza} and Landau\cite{LandauPR41}, that set the tone for the fundamental advances in the understanding of quantum macroscopic systems, and continuing with the most solid theoretical advances by Bogolubov\cite{Bogolubov} and Bardeen-Cooper-Schriefer (BCS) theories\cite{BCS,LeggettPRL72} of interacting Bose and Fermi gases, a picture has emerged in which it is clear that at very low temperatures and even moderate pressures, certain fluids undergo a phase transition to a superfluid phase. This phase is envisioned as a fluid with two parts, one a ``normal" fluid and another, a ``superfluid" fraction that does not contribute to the entropy\cite{Tisza,LandauPR41}. At zero temperature the system reaches its ground state but remains fluid, being superfluid all of it. The recent observation of quantized vortices in alkaline vapours of both Bose\cite{Matthews,Madison,Abo-Shaeer,Fetter-vortex} and Fermi\cite{Zwierlein} atoms, also confirms the superfluid nature of those gases and, hence, of the attained quantum macroscopic phase.

Although a satisfactory understanding of the nature of the corresponding quantum states near and at zero temperature has been made, as mentioned above, and a lot is known regarding general thermodynamic aspects of these fluids, see Refs. \cite{LondonBook,Grilly,Holian,Hoffer,Driessen,Huang} for instance, we would to point out here an aspect that seems to have gone unnoticed and this is the role of the chemical potential as an indicator of the transition to a superfluid state, related to the main idea that the superfluid fraction of the fluid does not contribute to the entropy. Despite all efforts to portray the appearance of superfluidity as a condensation phenomenon \`a la Bose-Einstein Condensation (BEC), we shall point out a striking property of {\it ideal} Fermi gases that serves as an illustration of the main hypothesis here advanced, namely, that the {\it condensation} phenomenon is the appearance of the {\it many-body ground state of the particles in the condensate}, and not of the occupation of a {\it single} particle state. We insist right away that this hypothesis is independent of Bose or Fermi statistics, but certainly, ideal BEC is included by this identification. Of course, the quantitative details of behavior of different fluids may strongly depend on the statistics and other peculiarities of the atoms or molecules in question, but no so their overall thermodynamic behavior leading to a superfluid phase.

The present article is based on a thermodynamic approach rather than one based on many body quantum mechanics. As a matter of fact, we lack such a theory to fully back the hypothesis put forward here, and a hope is that these ideas may show the way for finding it. In the last sections we point out that the present experiments with trapped alkaline gases may be interpreted in such a way as to being a direct measurement of the chemical potential and, thus, may lead to a corroboration of some aspects here discussed.

We proceed as follows. Section II is a brief review of general thermodynamic results useful for the following sections. Section III is devoted to show the relevant role played by the chemical potential. It is first argued that one can define it in an unambiguous way. Then, using its main definition, as being the negative of the change of entropy as a function of varying particle density at constant energy, we state the main hypothesis that negative chemical potentials correspond to normal phases while positive ones to quantum phases, zero value being the onset of the transition. A quantum phase is identified as that one in which  there appears a {\it condensate}, as defined above, that does not contribute to the entropy. These phases, for interacting fluids, show the phenomenon of superfluidity. In this same section we discuss ideal Bose and Fermi gases and show that, besides the common ideal BEC, there clearly appears a phenomenon that may be called Fermi-Dirac Condensation. In Section IV, based mainly in experimental phase diagrams of $^3$He and $^4$He  we attempt to build a qualitative ``generic" phase diagram that obeys all thermodynamic requirements. The main result is the phase diagram entropy density versus particle density, where one can plot {\it isoenergetic} curves, exemplifying our concerns. We advance the fact that the transition gas to superfluid in Helium and alkaline vapours is remarkably closer to that of ideal Fermi-Dirac condensation. Also, by continuity, we find that there should be ``supersolid" phases, in the sense that their chemical potential is positive, although we do not claim them to be the same as those supersolid phases currently discussed\cite{Chan,BalibarSS}. As mentioned, Section V is devoted to discuss the results of recent experiments in ultracold gases, in particular, we argue that the density profiles measured open the door for a full determination of the phase diagram, for both uniform and trapped regions. We conclude with some remarks. 

\section{Brief review of Thermodynamics}

For a pure (non-magnetic and non-polarizable) substance and confined by a vessel of rigid walls of volume $V$, all thermodynamics may be obtained from the knowledge of the entropy $S$ as a function of the internal energy $E$, the number of atoms or molecules $N$ and the volume $V$, namely, from the relationship $S = S(E,V,N)$. Since $S$ is extensive, one can write
\begin{equation}
S = V s(e,n) \label{F1}
\end{equation}
where $s = S/V$, $e = E/V$ and $n = N/V$ are respectively, the entropy, energy and number of particles per unit of volume. Thermodynamics is obtained from the following expressions that summarize the First Law,
\begin{equation}
ds = \beta de \> - \> \alpha dn, \label{F2}
\end{equation}
where the inverse temperature $\beta = 1/k_BT$ and chemical potential $\alpha = \mu/k_BT$ are,
\begin{equation}
\beta = \left(\frac{\partial s}{\partial e}\right)_n \>\>\>\>\>\>{\rm and}  \>\>\>\>\>\> \alpha = - \left(\frac{\partial s}{\partial e}\right)_e ,\label{C1}
\end{equation}
with  $k_B$ Boltzmann constant. We note that $\alpha$ is a dimensionless variable that we shall refer to both this and $\mu$ as the chemical potential. The pressure $p$ of the system is obtained via Euler expression, that is to say,
\begin{equation}
\beta p = s - \beta e + \alpha n .\label{F3}
\end{equation}
Thermodynamics assumes that $s$, $\beta$, $\alpha$ and $p$ are single valued functions of $e$ and $n$. The Second Law states that $s = s(e,n)$ is a concave function of its variables $(e,n)$, namely,
\begin{equation}
\frac{\partial^2 s}{\partial e^2} < 0 \>\>\>\>\>\>{\rm and}  \>\>\>\>\>\> \frac{\partial^2 s}{\partial n^2} - \frac{\left(\frac{\partial^2 s}{\partial n \partial e}\right)^2}{\frac{\partial^2 s}{\partial e^2}} < 0 .\label{C2}
\end{equation}
These conditions assert the stability of the thermodynamic states, the first inequality implies the positivity of the specific heat at constant volume $c_v$ while the second one the positivity of the isothermal compressibility, 
\begin{equation}
c_v = - \beta^2 \left( \frac{\partial e}{\partial \beta} \right)_n > 0 \>\>\>\>\>\>{\rm and}  \>\>\>\>\>\> \kappa_T = \frac{\beta}{n^2}  \left( \frac{\partial n}{\partial \alpha} \right)_\beta > 0 \label{C3}
\end{equation}
The second expression is equivalent to the more common one $\kappa_T = (1/n) (\partial n/\partial p)_\beta$. 

The Third Law asserts that Absolute Zero, $\beta \to \infty$, is unattainable and, if we consider systems with energy spectra unbounded from above, one concludes that the temperature is always positive, $\beta \ge 0$\cite{Ramsey}. The Laws are silent regarding the sign of the chemical potential $\alpha$. We shall argue that there is a profound meaning in the sign of this quantity, typically being negative, while its becoming positive signals the appearance of ``quantum" fluid phases. But before getting into that aspect, let us review briefly the characteristics of first and second order phase transitions.

Phase transitions only occur in the thermodynamic limit, where $S$, $E$, $V$, and $N$ become infinitely large, while the densities $s$, $e$, and $n$ remain finite. At a first order phase transition the (originally) extensive quantities $s$, $e$ and $n$ become discontinuous while the intensive ones, $\beta$, $\alpha$ and $p$ remain continuous. A phase transition of the second order occurs with a continuous change of all variables, $s$, $e$, $n$, $\beta$, $\alpha$ and $p$, however, the state is strictly unstable in the sense that the stability conditions of the Second Law, Eqs.(\ref{C2}), are no longer negative but become zero. That is, $c_v^{-1} \to 0$ and $\kappa_T^{-1} \to 0$. Or, equivalently, $c_v \to \infty$ and $\kappa_T \to \infty$ at the transition. By well known relationships of statistical physics\cite{LandauI}, the former indicates a divergence of the density energy fluctuations, while the latter a divergence of the particle density fluctuations. Thus, in the strict thermodynamic limit, a system cannot be stabilized at a second order phase transition. For our purposes below, a second order phase transition implies that both the pressure $p$ and the chemical potential $\alpha$, as functions of $n$ for a given value of the temperature $\beta$, become flat at the transition.

With the previous information plus empirical data regarding the overall characteristics of the phase diagram of the equation of state $p = p(n,\beta)$ one can construct the structure of the fundamental relationship $s = s(e,n)$. We stress out the important fact that the equation of state $p = p(n,\beta)$  is not a fundamental relationship in the sense that not all the thermodynamics can be obtained from its knowledge. One needs an additional function, such as $\alpha = \alpha(n,\beta)$. However,  $s = s(e,n)$ is fundamental since it does contain all thermodynamics.

\section{The chemical potential}

\subsection{The non-ambiguity of the chemical potential}

As mentioned above, the sign of the chemical potential $\alpha$ is not restricted by the Laws of Thermodynamics. Moreover, there is a widespread belief that the chemical potential is defined up to an arbitrary additive constant. Thus, enquiring about its sign and the meaning of it, may appear as a futile exercise. We now argue that such a belief is not correct and establish that the sign of the chemical potential is of fundamental relevance.

An alternative determination of the chemical potential to that of Eq.(\ref{C1}) is the following,
\begin{equation}
\beta \alpha = \left(\frac{\partial e}{\partial n}\right)_e .\label{cp2}
\end{equation}
This identity follows from the uniqueness of $s$ as a function of $e$ and $n$. The argument of the indeterminacy of $\alpha$ is that since the system is non-relativistic, the energy $E$ can be defined up to an arbitrary constant $C$ thus shifting its origin arbitrarily, $E^\prime = E + C$, and because of Eq.(\ref{cp2}), yielding a chemical potential relative to such an arbitray origin of the energy. Let us review this carefully. First, since the thermodynamic limit must be imposed, clearly, an arbitrary constant $C$ would drop out since $e^\prime = E^\prime/V + C/V \to E/V = e$. Thus, the arbitrary constant should be extensive, that is, $C = N c$, with $c$ indeed an arbitrary constant. This would certainly imply $\beta \alpha^\prime = \beta \alpha + c$. We now argue that the constant $c$ can always be (as it is done in practice) set equal to zero. 

Consider a realistic model of a pure fluid. It is first assumed to be non-relativistic. Thus, the Hamiltonian of $N$ atoms (or molecules), classical or quantum,  may be written as,
\begin{equation}
H = \sum_{i=1}^N \frac{\vec p_i^2}{2m} + V_{int}(\vec r_1, \vec r_2, \dots, \vec r_N) + \sum_{j=1}^N V_{ext}(r_j) ,\label{H1}
\end{equation}
where $V_{ext}(\vec r)$ is the common confining potential, and here we are concerned with rigid-walls potentials only. The interatomic potential is $V_{int}(\vec r_1, \vec r_2, \dots, \vec r_N)$ and can contain two-, three-, etc., body interactions. This potential can be defined, again, up to an arbitrary constant. However, for the energy $E$ to be extensive, it should be true that in the limit of ``infinite dilution", namely, in the limit $|\vec r_i - \vec r_j| \to \infty$ for all pairs of particles $i$ and $j$ in the system, the potential should take the form $V_{int}(\vec r_1, \vec r_2, \dots, \vec r_N) \to Nc$. In all textbooks is assumed that the arbitrary constant is zero, $c \equiv 0$. That is, one always tacitly assume that in such a limit the system behaves as an {\it ideal} gas (quantum or classical). We shall take this point of view here. This means that the energy is uniquely defined and so does the chemical potential $\alpha$. With regard to this common convention, we can establish the meaning of the sign of $\alpha$.

\subsection{The sign of the chemical potential}

Recall the definition of the chemical potential, Eq.(\ref{C1}). It gives the negative of the change of entropy with respect to a change of particles, for a given value of the energy. Thus, if negative (positive), it tells us that the entropy increases (decreases) with increasing number of particles, for a fixed value of the energy. We now resort to the well-known recipe from statistical physics\cite{LandauI} to calculate the entropy of an open system in interaction with a heat bath, in terms of its (average) energy $E$, number of particles $N$ and volume $V$,
\begin{equation}
S(E,N,V) = k_B \ln \Delta \Gamma(E,N,V) ,\label{S1}
\end{equation}
where $\Delta \Gamma(E,N,V)$ is the number of states the system has access when in thermal equilibrium with its environment. This number of states is defined within an interval $\Delta E$, the energy fluctuation of the thermodynamic state, namely,
\begin{equation}
\Delta \Gamma = \frac{d \Gamma}{d E} \> \Delta E,
\end{equation}
where $d \Gamma/d E$ is the density of states. For a fixed value of the energy $E$ and the volume $V$, the statistical weight $\Delta \Gamma(N)$ is a function of the number of particles.

To elucidate the dependence of $s$ on $n$, let us consider a gas at large enough temperatures such that the system may well be approximated as classical and even close to an ideal gas. Let us keep the energy $e$ fixed at that corresponding value. Then, it should be clear, and textbook experience corroborates it, that if the number of particles is increased the number of available states does too increase. That is, there is an increase of available states by enlarging the phase space with more particles. The chemical potential $\alpha$ is thus negative. One finds, however, that the energy per particle $\epsilon = e/n$ obviously decreases and, at the same time, not completely obvious but corroborated by simple cases and verified below, the temperature $T$ also {\it decreases}. It is important to realize that as the number of particles is increased, the structure of the states and of the energy spectrum do change, that is, the system is described by a different Hamiltonian, since this depends on $N$. Consider now that the energy is still kept constant but the number of particles keeps increasing even further. Two situations may occur:

i) The system reaches a {\it limiting} state of finite temperature. That is, the system reaches a state where if the number of particles is increased even more, the energy can no longer be kept constant. That is, the system must suffer a first order phase transition, the temperature, pressure and chemical potential changing continuously, but the energy and entropy becoming discontinuous, typically increasing. In brief, the isoenergetic curves eventually meet a state that borders an unstable region, where phase separation occurs, but the important point is that the chemical potential is always negative $\alpha < 0$ since, in this case, entropy is an increasing function of $n$ . We call ``normal" to these states.

ii) The system reaches the {\it limiting} state of zero temperature. This case may occur because the energy spectrum of the many-body system is bounded from below by the ground state energy $E_0 = E_0(N,V)$. Thus, for a fixed value of the energy $e$, there exists a {\it maximum} value of the density $n_{max}$ at which the energy of the ground state of that density equals the energy $e$, namely, $e_0(n_{max})=e$, where $e_0 = E_0/V$. At this stage the system is in its {\it corresponding} ground state, the entropy becomes zero $s(e,n_{max}) = 0$ and so does the temperature $T = 0$. Since at low enough density the entropy increases with increasing density but eventually reaches $s = 0$ for $n_{max}$, it follows that the entropy must have reached a maximum value at some critical density $n^*$, above which the entropy begins to decrease. This in turn implies that the chemical potential is negative for small enough density, becoming zero at $n^* < n_{max}$ and turning into a positive quantitive above it. We shall argue further below that this change signals the appearance of, or transition to, a quantum fluid or a quantum solid with proper macroscopic properties. Let us advance a hypothesis of the nature of that state that explains the decrease of the entropy at the point $\alpha = 0$.

\subsection{The nucleation of a condensate phase}

As mentioned above, if for a fixed value of the energy the system reaches a limiting state at zero temperature, which in turn is the ground state for that density, it must be true that in the vicinity of that state the chemical potential must be positive indicating that the entropy {\it decreases} with increasing density. This implies that at some density $n^*$ the chemical potential becomes zero $\alpha(e,n^*) = 0$. One can conclude that as this point is crossed by increasing the density, a macroscopic phase nucleates such that it does not contribute to the entropy. To have such a property, this phase must be a {\it single} many-body quantum state. As we shall analyze below, this coincides with the appearance of superfluid phases in He and alkaline ultracold vapors.  For a Bose ideal gas, this is also the phenomenon of Bose-Einstein Condensation (BEC). What we further claim here, as the main hypothesis of this study, is that this phase must be very close to the many-body ground state of the fraction of the particles that  conforms such a phase. Let $n_c$ be the fraction of particles (per unit volume) in the zero-entropy phase and $n_T= n - n_c$  the remaining ones. Thus, we claim, the zero-entropy phase is the many-body ground state of the $n_c$ particles with energy $e_0(n_c)$, i.e. $e(n) = e_T(n_T) + e_0(n_c)$. As the number of particles is increased, $n_c$ increases up to the point where $n_T = 0$ and $n_c = n_{max}$, and the whole system is in its corresponding ground state. The entropy and the temperature are zero. We immediately point out that this {\it is not} the usual Bose-Einstein condensation scenario. In such a case the condensation occurs at a {\it single}-particle state. Here, we insist in the fact that the condensate is the {\it many-body} ground state of the condensate particles. Of course, the many-body ground state of $n_c$ particles per unit volume in the ideal Bose gas is the same as  $n_c$ particles per unit volume in the ground state of {\it one} particle. We discuss now that the most common case of condensation of a  {\it many-body} ground state is already present in the ideal Fermi gas and that this is actually closer to the scenario of real superfluids.

\subsection{Ideal Bose-Einstein and Fermi-Dirac Condensation}

\begin{figure}
  \begin{center} \scalebox{0.75}
   {\includegraphics[width=\columnwidth,keepaspectratio]{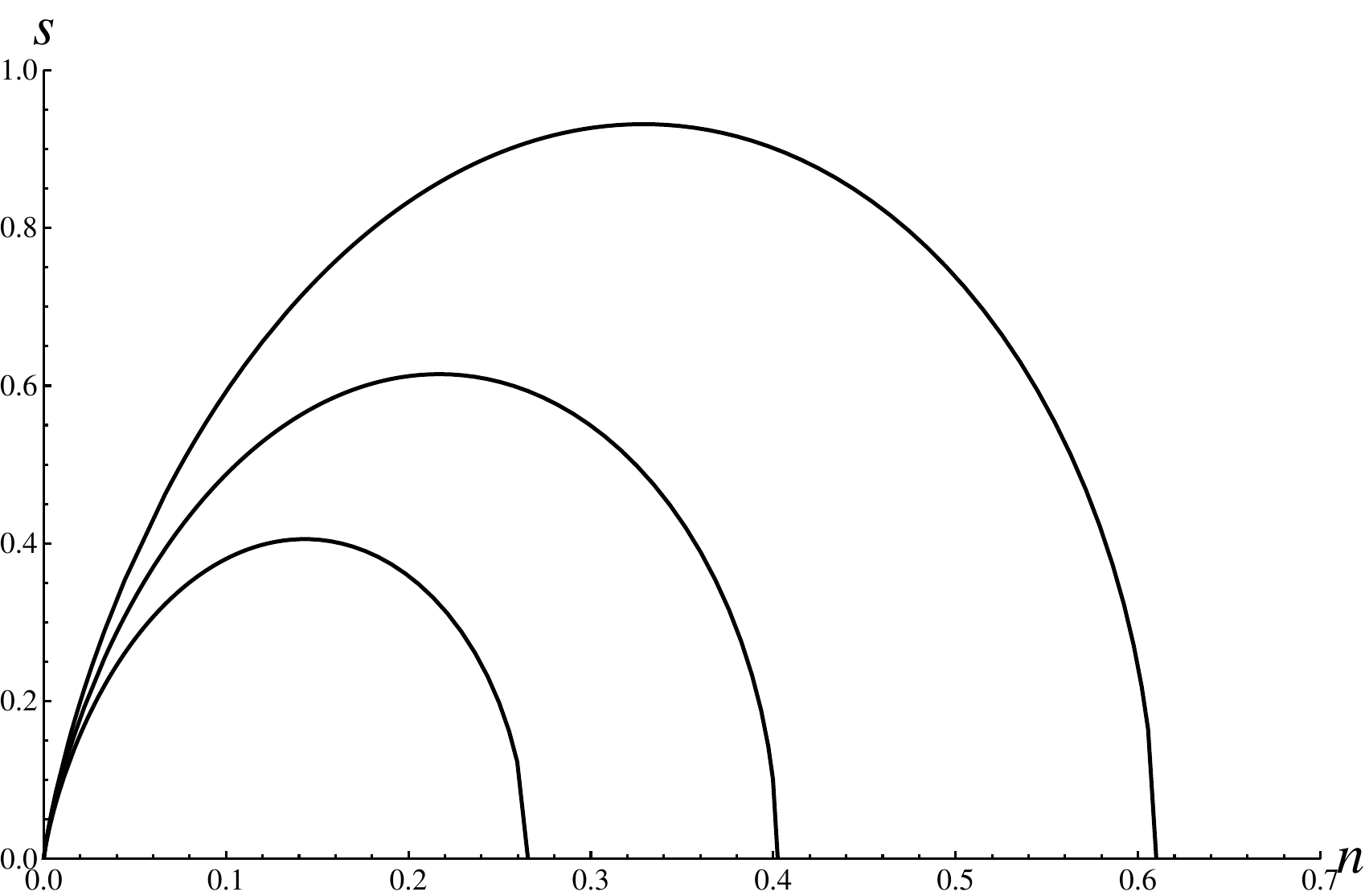}}
   \caption{Entropy $s$ versus density $n$ at a fixed value of the energy $e = 0.5, 1.0$ and 2.0 for an ideal Fermi  gas, in units $\hbar = m = k_B = 1$. The slope of the curve is the negative of $\alpha = \beta \mu$.}\label{fig3A}
  \end{center}
\end{figure}

\begin{figure}
  \begin{center} \scalebox{0.75}
   {\includegraphics[width=\columnwidth,keepaspectratio]{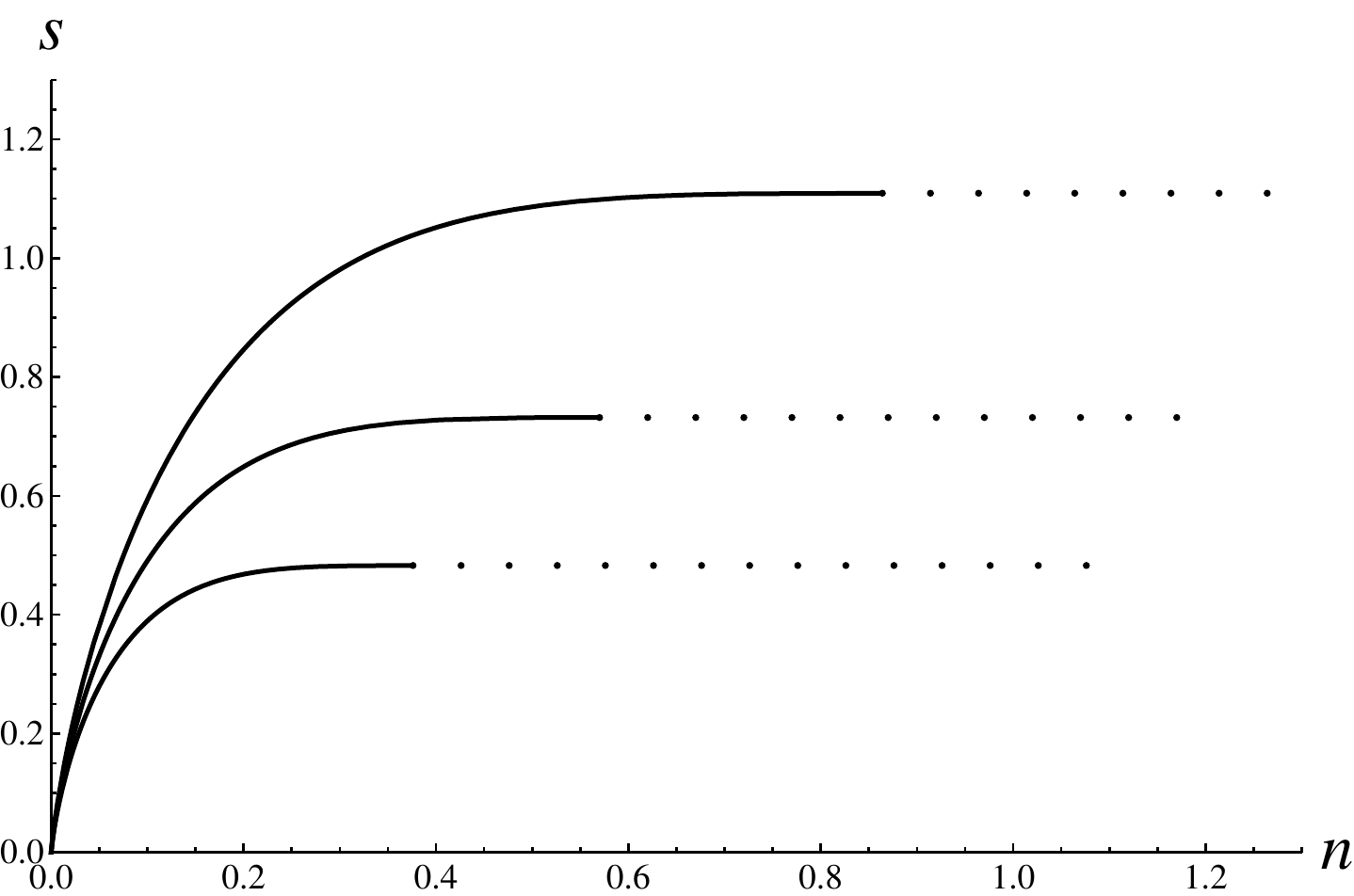}}
   \caption{Entropy $s$ versus density $n$ at a fixed value of the energy $e = 0.5, 1.0$ and 2.0 for an ideal  Bose gas, in units $\hbar = m = k_B = 1$. The slope of the curve is the negative of $\alpha = \beta \mu$. The dotted lines correspond to Bose-Einstein condensation states. }\label{fig3B}
  \end{center}
\end{figure}

It is common knowledge that Bose-Einstein condensation in ideal gases is a phenomenon characteristic of Bose statistics only. Here we argue that as a matter of fact, the phenomenon of condensation, understood as the development of a quantum phase that decreases the entropy of the system and characterized by a positive chemical potential, is also present in ideal Fermi-Dirac gases. In Figs. \ref{fig3A} and \ref{fig3B} we show the entropy $s$ as a function of the density $n$ for fixed values of the energy $e$, for ideal gases\cite{aclara}. Fig.  \ref{fig3A} is for Fermi and Fig.  \ref{fig3B} for Bose statistics. Both curves begin, at very low density, with chemical potential negative. At some density $n^*$ both curves reach a maximum and the chemical potential $\alpha$ becomes zero. For bosons this is the onset of the well known phenomenon of Bose-Einstein condensation, and beyond this point the chemical potential remains zero. The interpretation is that the {\it single particle} ground state begins to be macroscopically occupied. That is, at fixed energy, for any given value of the density $n >n^*$, there is a fraction $n_c = n -n^*$ that occupies the single-particle ground state. We note, however, that such a state is also the many body ground state of the $n_c$ particles. Moreover, the energy of the system equals that of the critical density $e = e(n^*)$ since the condensate has zero energy. For fermions, the zero chemical potential point is essentially ignored in most textbook discussions. We argue now that this also marks the onset of a phenomenon that may also be called Fermi-Dirac condensation. The condensation is not the macroscopic occupation of a single-particle state but rather of a many-body ground state, that of the condensed fraction. 

\begin{figure}
  \begin{center} \scalebox{0.75}
   {\includegraphics[width=\columnwidth,keepaspectratio]{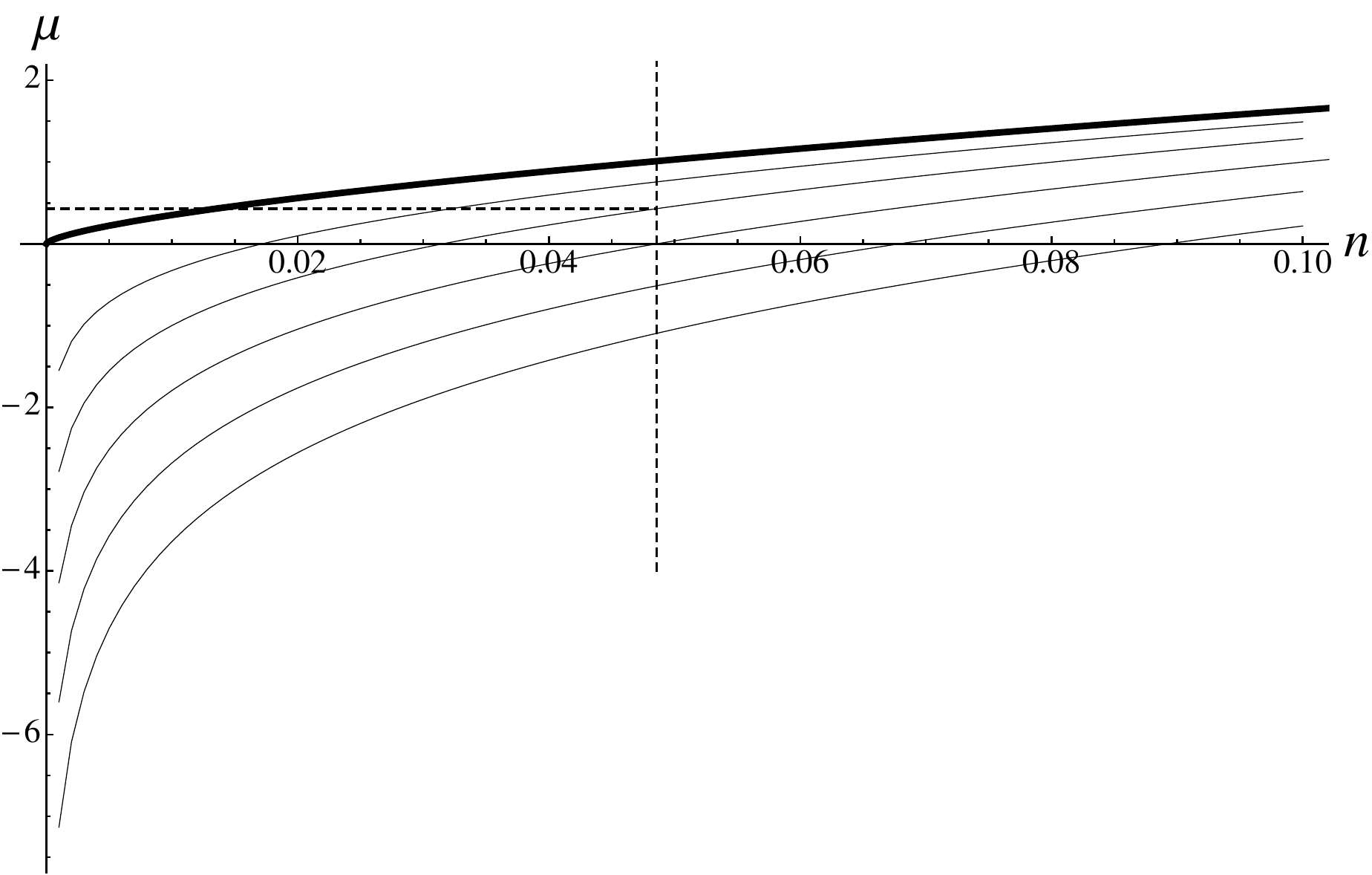}}
   \caption{Isotherms ($T$ = const) $\mu$ vs $n$ for an ideal Fermi-Dirac gas. The thick curve is $T = 0$ (Fermi energy). The vertical dashed line indicates an arbitrary density $n$. This line crosses all isotherms and defines the critical one $T_c(n)$ at $\mu = 0$. For temperatures $T < T_c(n)$ the density separates into a condensate fraction $n_0$ and a thermal one $n_T$:  at the temperature $T$ defined by the crossing of the vertical and horizontal dashed lines, the condensate fraction $n_0(T)$is the intersection of the horizontal dashed line with the $T = 0$ curve, while the thermal part is $n_T(T) = n - n_0(T)$. Clearly, $n_0$ becomes zero for $T > T_c(n)$ and $n_0 = n$ at $T = 0$.}\label{muvsnFD}
  \end{center}
\end{figure}

\begin{figure}
  \begin{center} \scalebox{0.75}
   {\includegraphics[width=\columnwidth,keepaspectratio]{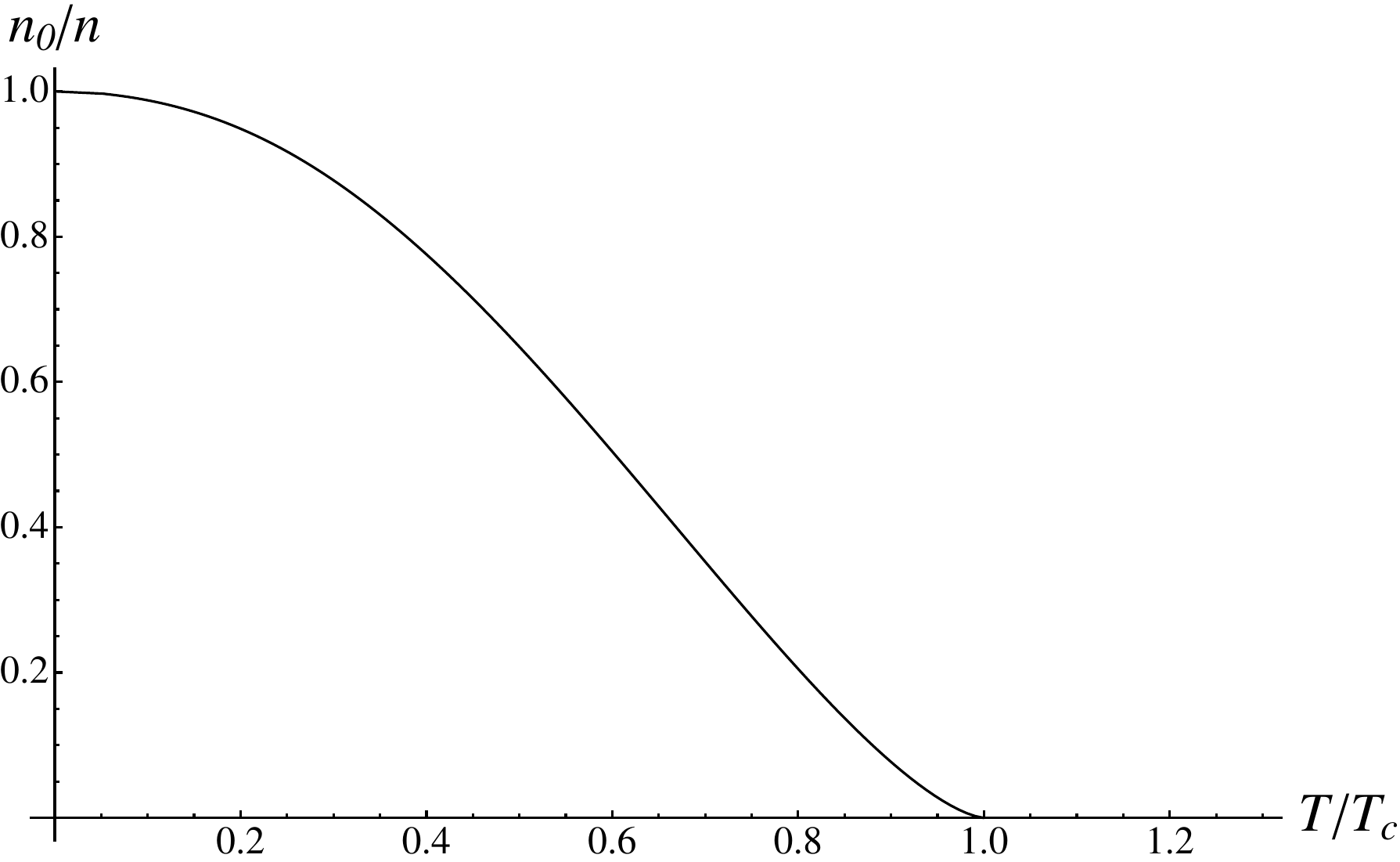}}
   \caption{Fermi-Dirac Condensation. Plot of $n_0/n$ vs $T/T_c$ where $n_0(n,T)$ is the fraction of particles that occupy its {\it own} many body ground state. $T_c(n)$ is the critical temperature for density $n$ obtained when the chemical potential becomes zero, see Fig. \ref{muvsnFD}.}\label{FDcond}
  \end{center}
\end{figure}

From  Fig. \ref{fig3A} we see that, at fixed energy, the entropy reaches a maximum at a density value $n^*$ where the chemical potential becomes zero. Beyond this point the chemical potential becomes positive indicating a decrease of the entropy.  What does occur in the gas? A macroscopic occupation of a single-particle state is forbidden by Fermi statistics. However, we argue, there appears a ``condensate" fraction which is in a {\it single many body} state in order to give a zero contribution to the entropy. This condensate grows a $n$ is increased thus decreasing the entropy. 
This behavior continues up to a maximum value $n_{max}$ where the entropy is zero and where {\it all} the gas is in its corresponding many body ground state. The latter state has an energy $e_0(n_{max})$. Note that at this point $\alpha \to \infty$, $\beta \to \infty$ but the ratio $\mu = \alpha/\beta$ equals the so-called Fermi energy $\epsilon_F$. We argue then, that for values between $\mu = 0$ and $\mu = \epsilon_F$, a fraction $n_c$ must occupy a single many-body state that does not contribute to the entropy. By continuity in the neighborhood of zero temperature, that is, at $n_{max}$ where all the gas is indeed in its many body ground state, the condensate must also be in its {\it own} many body ground state. To find this condensate fraction we appeal to Fig. \ref{muvsnFD} where we plot $\mu$ vs $n$ for different values of the temperature. The thick line denotes the value of the chemical potential at zero temperature $\mu_0(n) \equiv \mu(n,T=0)$, i.e. $ \mu_0(n) = \epsilon_F(n)$. Take a {\it fixed} value of $n$, say the vertical dotted line in Fig \ref{muvsnFD}. Let $T_c (n)$ be the temperature at which the chemical potential is zero. For temperatures $T < T_ c(n)$, the chemical potential is positive, $\mu > 0$, and the total fixed density equals $n = n_T + n_0(\mu)$, where $n_0(\mu)$ is the value of the density obeying $\mu_0(n_0) = \mu$. This is indicated by the horizontal dotted line in Fig. \ref{muvsnFD}. The thermal density $n_T$ is obtained from the difference of $n - n_0(\mu)$. For temperatures $T > T_c(n)$, clearly $n_0(\mu)=0$, all the gas is in a thermal state and the chemical potential is negative, $\mu < 0$.
Figure \ref{FDcond}  is a plot of $n_0/n$ vs $T/T_c(n)$ obtained from Fig. \ref{muvsnFD} for any value of $n$, that is, it is a universal curve. In analogy to Bose-Einstein condensation, this phenomenon may be called Fermi-Dirac condensation for particles obeying Fermi statistics. We repeat once more, the condensate is the fraction of the density that is in its {\it own} many-body ground state. As explained above this state coincides with the one-particle ground state condensate for ideal Bose gases. In the next sections we shall present a generic phase diagram for real interacting quantum fluids that present the phenomenon of superfluidity. We will see that such a state behaves more like the ideal Fermi-Dirac condensation than the ideal Bose-Einstein one.

\section{Generic phase diagrams of substances that show quantum phases}

As mentioned in Section II, the relationship $s = s(e,n)$ is a fundamental one in the sense that it contains all the thermodynamic information of the system. To construct it we may follow the recipes of statistical mechanics\cite{LandauI} and/or resort to empirical evidence. The latter is typically obtained with the measurement of the equation of state $p = p(n,T)$. In this section, based on the empirical equations of state of $^3$He and $^4$He\cite{London,basic,Fin}, and using the general properties of the equilibrium states as describe in Section II, we construct  {\it generic} phase diagrams of $s = s(e,n)$. In the next section, we argue that many of the characteristics of those phase diagrams should remain valid for bosons or fermions systems whether confined or not.

\begin{figure}
  \begin{center} \scalebox{1.0}
   {\includegraphics[width=\columnwidth,keepaspectratio]{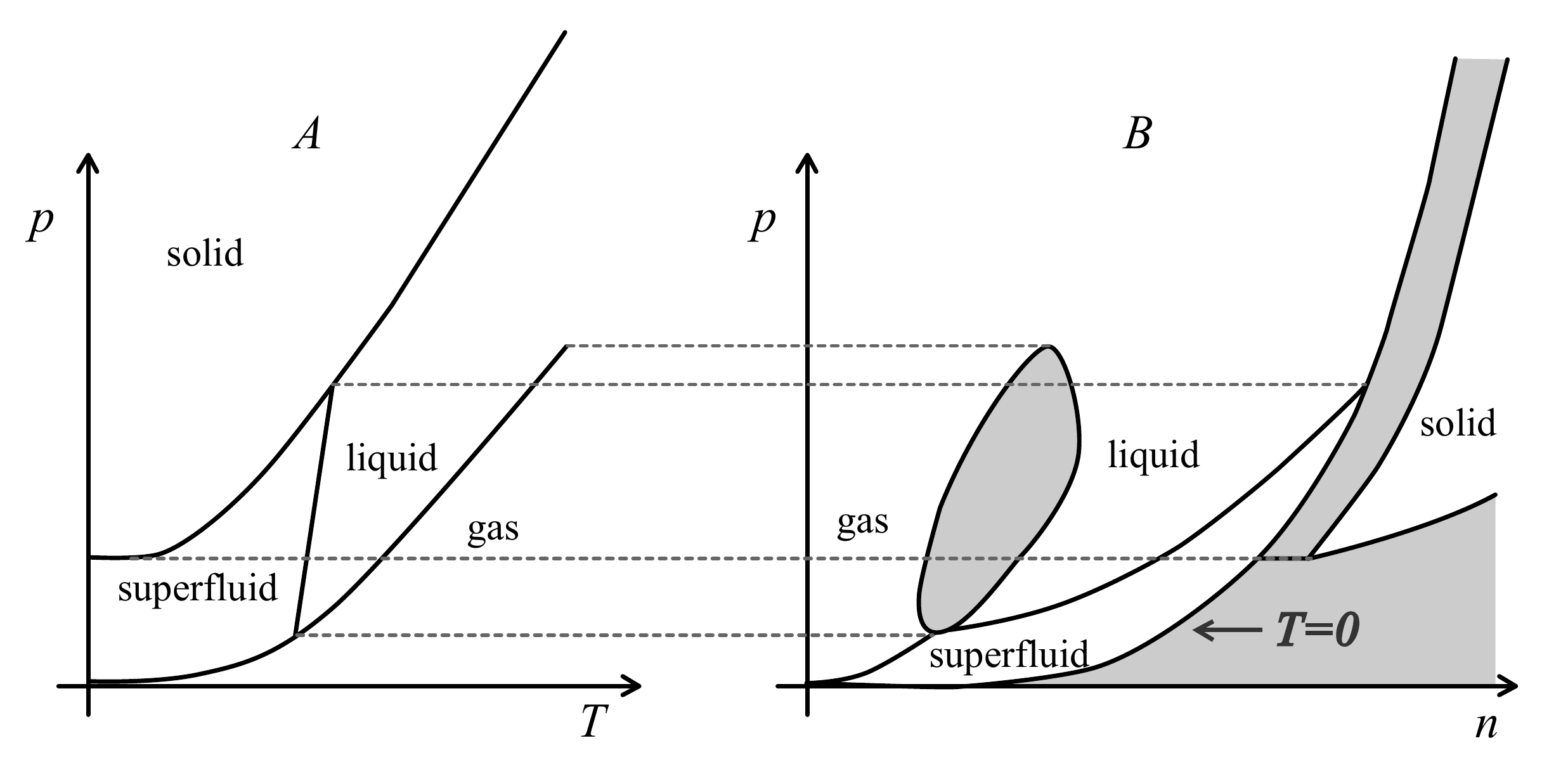}}
   \caption{Qualitative phase diagrams of a pure substance showing a superfluid phase. The areas in gray color are forbidden unstable regions.}\label{fig5}
  \end{center}
\end{figure}

Fig. \ref{fig5}A shows a phase diagram $p$ vs $T$ that qualitatively resembles that of $^4$He and $^3$He\cite{London,basic,Fin}. Four phases are present, gas, liquid, superfluid and solid. The liquid - solid,  liquid - gas  and superfluid - solid transitions are believed to be first order phase transitions, while liquid - superfluid and gas - superfluid second order (as well as the isolated critical point in the normal liquid - gas transition).  We consider all first order transitions being discontinuous in the particle, energy and entropy densities.  The transition line liquid to superfluid is supposed to have positive slope. This is true in $^3$He while it is negative in $^4$He .A transition line with negative slope is considered ``anomalous" since some properties do not follow ``expected" behavior near the transition. However, there is  no fundamental requirement for the sign of coexistence lines and we have assumed all of them to be positive for simplicity. We mention that even at zero magnetic field the $^3$He phase diagram shows two superfluid phases (so called $A$ and $B$)\cite{Osheroff,Grilly} due to the particular magnetic properties of  $^3$He. Again, for simplicity, we consider our hypothetical system to show the simplest of phase diagram, including nevertheless, a superfluid phase

\begin{figure}
  \begin{center} \scalebox{1.0}
   {\includegraphics[width=\columnwidth,keepaspectratio]{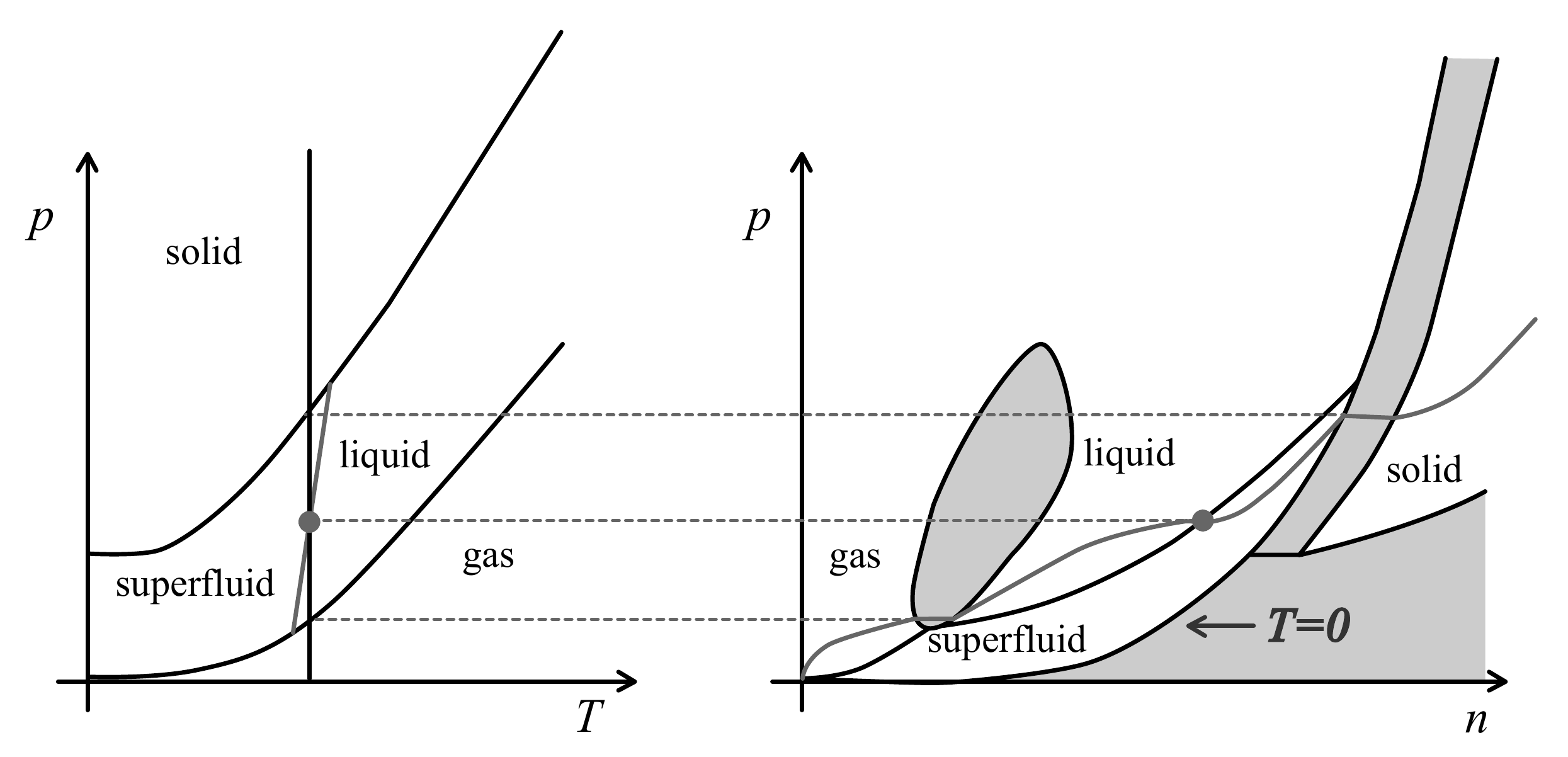}}
   \caption{Isotherm showing a gas-liquid first order, liquid-superfluid second order  and superfluid-solid first order phase transitions. By stability, see Eq.(\ref{C2}), the slope of the isotherm $p$ vs $n$ is always positive, except at a second order phase transition where it becomes flat (marked with a dot) showing the critical divergence of the compressibility. The areas in gray color are forbidden unstable regions.}\label{fig6}
  \end{center}
\end{figure}

\begin{figure}
  \begin{center} \scalebox{1.0}
   {\includegraphics[width=\columnwidth,keepaspectratio]{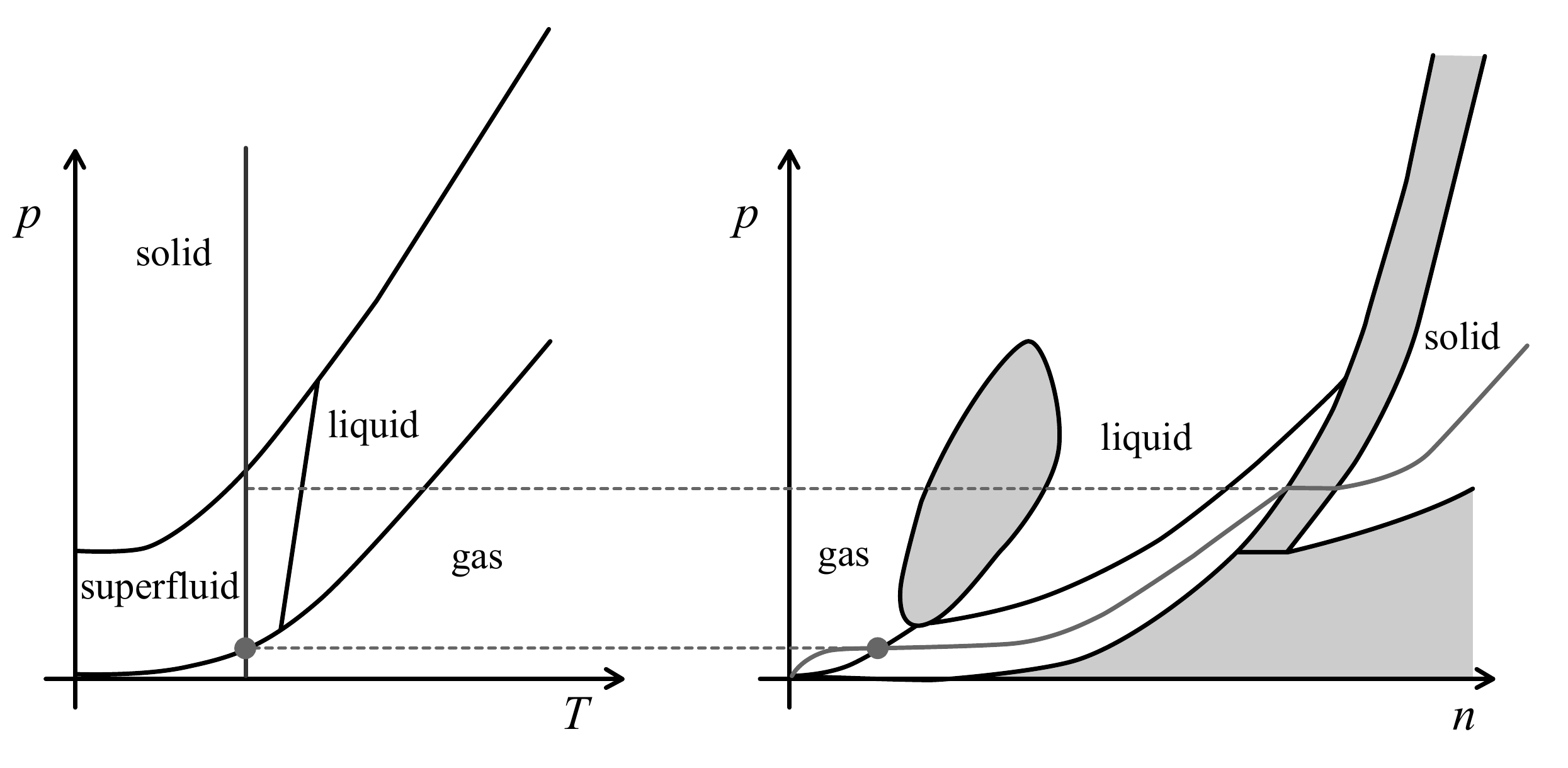}}
   \caption{Isotherm showing a gas-superfluid second order (marked with a dot) and a superfluid-solid first order phase transitions. The areas in gray color are forbidden unstable regions. See caption of Fig. 6.}\label{fig7}
  \end{center}
\end{figure}

Based on Fig. \ref{fig5}A and on general thermodynamic requirements one can construct the qualitative phase diagram $p$ vs $n$. This is shown in Fig. \ref{fig5}B. In Figs. \ref{fig6} and \ref{fig7} we show a couple of isotherms obeying the general requirements of Section II, indicating how the phase diagram $p$ vs $n$ of Fig. \ref{fig5}A was constructed. On the one hand, we appeal to the positivity of the isothermal compressibility, Eq.(\ref{C2}), to obtain $p$ as an increasing function of $n$ for $T$ constant. Second,  we observe that in first order phase transitions $n$ is discontinuous with $p$ continuous, while $p$ shows a flat slope at the density $n$ of second order transitions. 

\begin{figure}
  \begin{center} \scalebox{1.0}
   {\includegraphics[width=\columnwidth,keepaspectratio]{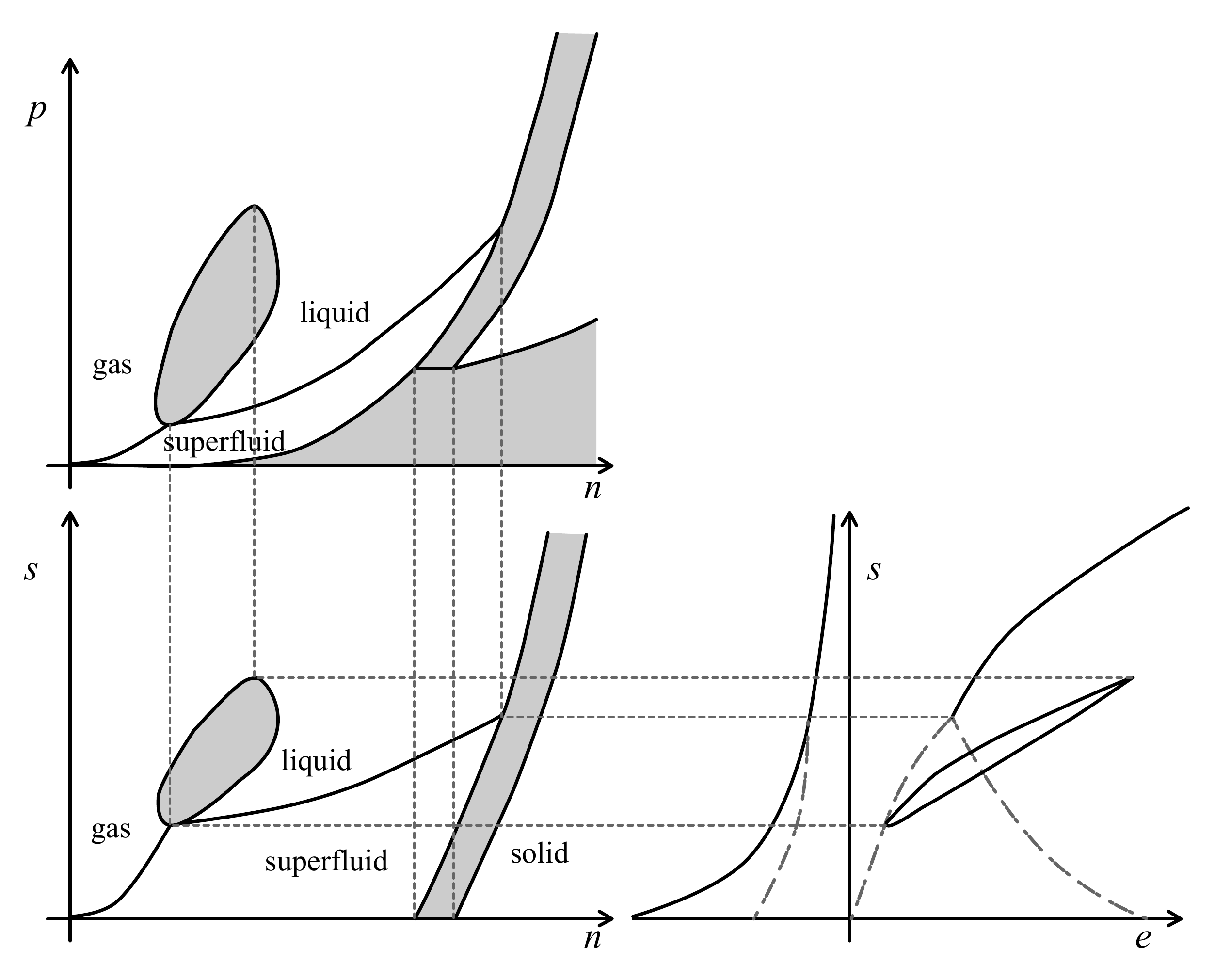}}
   \caption{Phase diagrams of the fundamental relationship $s = s(e,n)$. The areas in gray color are forbidden unstable regions.}\label{fig8}
  \end{center}
\end{figure}

\begin{figure}
  \begin{center} \scalebox{1.0}
   {\includegraphics[width=\columnwidth,keepaspectratio]{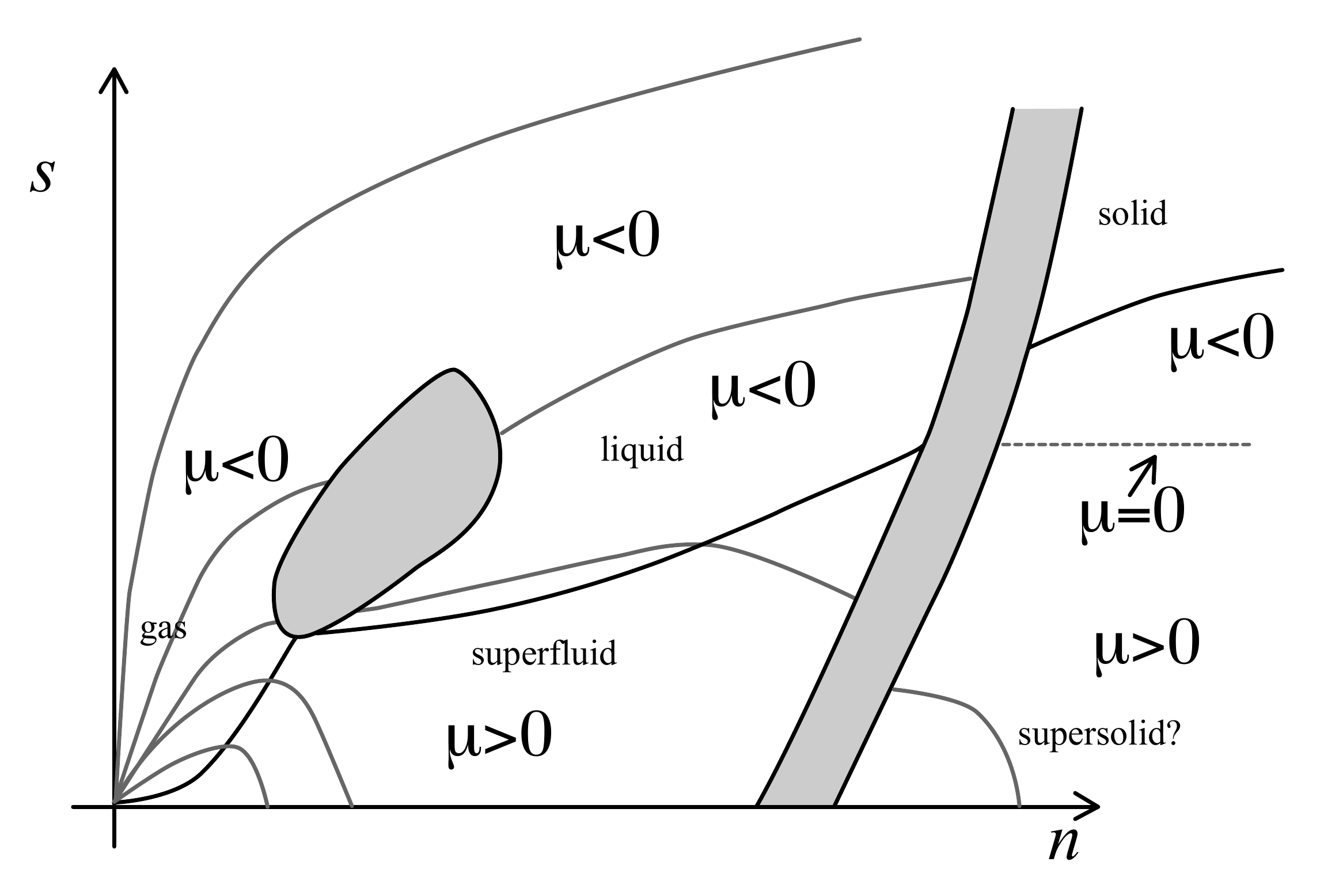}}
   \caption{Phase diagram $s$ vs $n$ showing isoenergetic curves indicating the regions of normal $\mu < 0$ and quantum $\mu > 0$ behavior. Stability requires those curves to have negative curvature. Its slope is the negative of $\alpha = \beta \mu$. We identify the normal to quantum transition as being characterized by $ \mu = 0$. }\label{fig9}
  \end{center}
\end{figure}

The relevance of the phase diagram $p$ vs $n$ shown in Fig. \ref{fig5} is that its allows us to qualitatively construct the phase diagram $s$ vs $n$. With this, and following the thermodynamic laws, one can then built $s$ vs $e$. These diagrams are shown in Fig. \ref{fig8}. These two diagrams embody all the thermodynamics of the system. For the purposes of the present article, we show in Fig. \ref{fig9} different {\it isoenergetic} lines $s$ vs $n$ whose meaning is discussed below. The discussion of the {\it isodensity} curves $s$ vs $e$ is not presented in this article since they can be very complicated and do not add more to the main point of this section.

Returning to Fig. \ref{fig9}, we show several isoenergetic lines $s$ vs $n$. By the stability of the states, see Eq.(\ref{C2}), their curvature must be negative and its slope is $\alpha = \beta \mu$. Since $\beta$ is always positive, this implies that the chemical potential has the same sign as $\alpha$. In Fig. \ref{fig9} we have indicated the regions of the phase diagram where $\mu$ is either positive or negative. We note that {\it a fortiori} there are regions where the chemical potential is negative and others where is positive. That it must be negative at some points is a consequence that in the appropriate limit the normal gas behaves as an ideal gas, while the positive sign region must exist because very near to $T = 0$  the system is essentially in its ground state and the entropy must be reduced as $n$ increases for $e$ fixed. We note the striking similarity of the isoenergetic curves in the gas to superfluid region of Fig. \ref{fig9} with that of the ideal Fermi gas in Fig. \ref{fig3A}. Our claim that this region marks the existence of a condensate fraction, being a many body ground state as explained above, is supported by the fact that the region $\mu  >0$ coincides with the superfluid macroscopic phase state. That is, it is well accepted that the superfluid {\it fraction} of the superfluid state at $T \ne 0$ does not contribute to the entropy. It appears natural to identify the condensate as the superfluid fraction. 

We would like to add that due to the continuity of the chemical potential (and temperature and pressure) at a first order phase transition, then, there must be {\it solid} phases with chemical potential negative and others with positive. The former are the ``normal" solids, while the latter indicate the existence of a ``quantum" solid or ``supersolid". The transition line between these two is within the solid and it should be a line of zero chemical potential. This is also indicated in the figure. We are not claiming that the ``supersolid" phase is necessarily the same as that recently discussed in the literature\cite{Chan,BalibarSS}. Further analysis of this phase is beyond the scope of this article.

To conclude this section, we point out that if we had started with a phase diagram $p = p(n,\beta)$ of a pure ``normal" substance, that is, one that does not show superfluid phases, one obtains that the chemical potential is always negative.

\section{Many body theories at $T = 0$ and recent experiments in ultracold confined gases}

As mentioned in the Introduction, many-body theories have predicted quite satisfactorily the ground state of interacting gases, thus describing the superfluid state at zero temperature. On the other hand, the details of the phase transitions to superfluid states is still a challenge to theoretical efforts. From experimental evidence is believed that transitions to superfluid from gas and liquid phases are second order. However, even at the level of mean field theories the description of those transitions is still a matter to be settled\cite{OQVRR1,OQVRR2}. In addition to the predictions of the celebrated theories mentioned above, we believe that the present experiments with ultracold gases show the opportunity to prove, if not inadvertendly already been done, that the phase diagrams presented above are the correct ones. In particular, the ultracold experiments have probed the region of very small temperatures and very small pressures, where the transition gas to superfluid occurs. In this section we address the phenomena observed in {\it trapped} inhomogenous ultracold gases, to argue that an analysis of experimental data can be made that should lead directly to the appropriate phase diagrams yielding further information on the order of the transition.

For our purposes we recall first the predictions of the fundamental theories regarding the behavior of the chemical potential as a function of the density. At zero temperature, both for fermions and bosons, the chemical potential is a {\it positive} increasing function of the density. That is, the chemical potential predicted by Bogolubov and BCS is very similar to that of an ideal Fermi gas (the thick line in Fig. \ref{muvsnFD}). We recall that Bogolubov prediction is $\mu(T = 0) \sim n$, while for BCS is essentially the ideal Fermi energy $\mu(T = 0) \sim n^{2/3}$. This result, in addition to the fact that superfluidity is present, seems to be in accord with the ideas presented in the previous sections. That is to say, at some point the chemical potential must have changed sign indicating the appearance of a condensate-superfluid phase.

\begin{figure}
  \begin{center} \scalebox{0.75}
   {\includegraphics[width=\columnwidth,keepaspectratio]{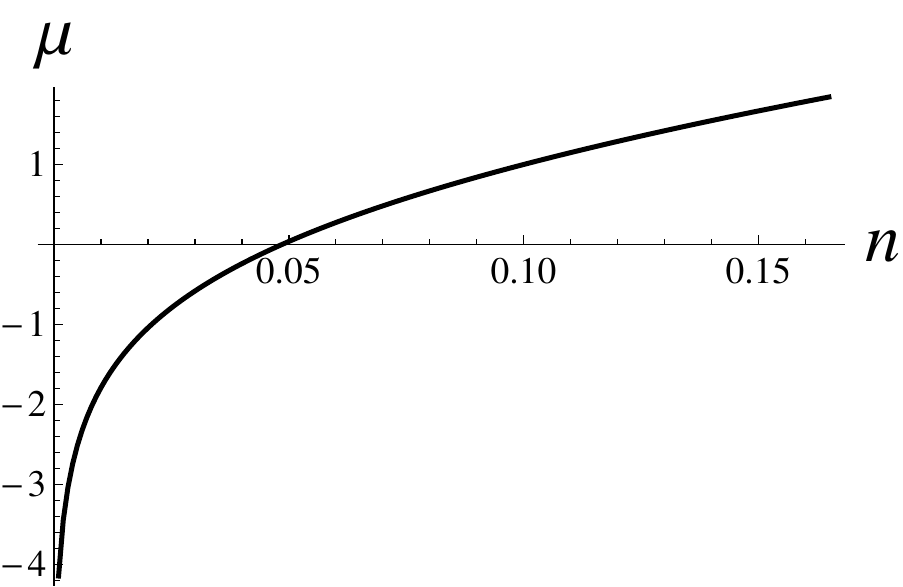}}
   \caption{Isotherm $n$ vs $\mu$ for $T = 1.0$, of an ideal Fermi gas. Units $\hbar = m= k_B = 1$. 
    }\label{nvsmu-FD}
  \end{center}
\end{figure}

\begin{figure}
  \begin{center} \scalebox{0.75}
   {\includegraphics[width=\columnwidth,keepaspectratio]{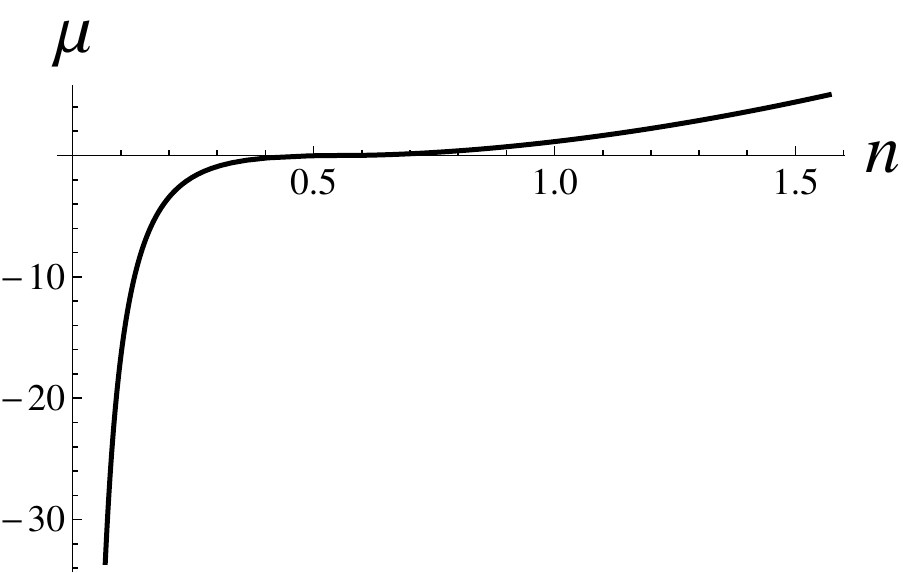}}
   \caption{Hypothetical isoterm $n$ vs $\mu$ of an intereracting quantum gas. As a matter of fact, for $\mu < 0$ we used an ideal Bose gas and a curve $\mu = C (n - n_c)^\gamma$ with $\gamma > 1$, for $\mu > 0$ where $C$ is a constant and $n_c$ is the ``critical" value of the density, namely, its value  at $\mu = 0$. The hypothesis is that the transition should be a second order phase transition and, therefore, the slope of $\mu$ vs $n$ should be zero at $\mu = 0$. }\label{muvsn-h}
  \end{center}
\end{figure}

\begin{figure}
  \begin{center} \scalebox{0.75}
   {\includegraphics[width=\columnwidth,keepaspectratio]{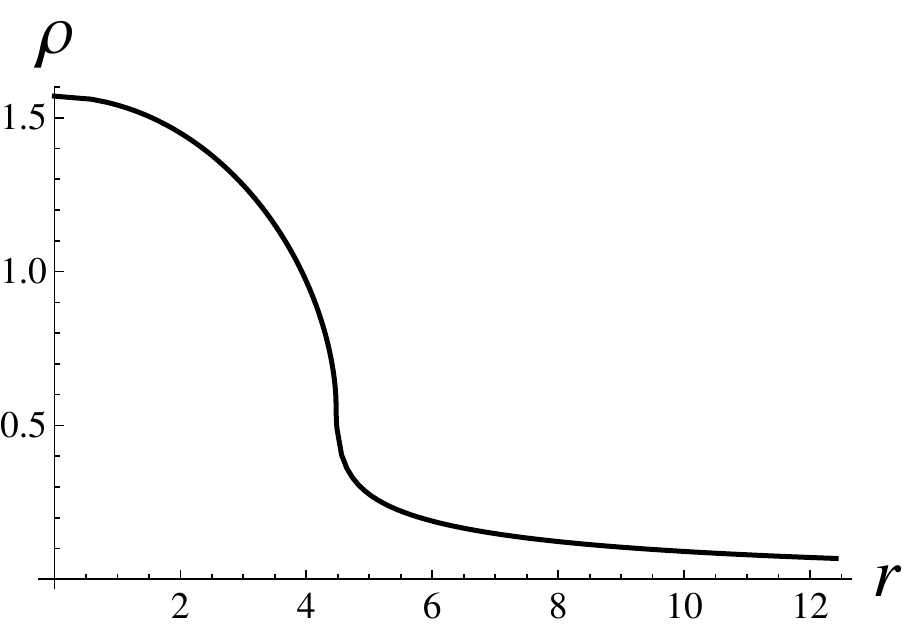}}
   \caption{Hypothetical density profile $\rho(r)$ vs $r$ of an interacting quantum gas confined by a harmonic trap, for a temperature below condensation $T < T_c$. This profile was obtained using the local density approximation, see Eq.(\ref{profile}), of the isotherm $\mu$ vs $n$ in Fig. \ref{muvsn-h}. The frequency is $\omega = 1$ in units of $\hbar= m = k_B = 1$. Typical experimental density profiles look like this, with a thermal gaussian tail and a peak indicating the presence of the condensate.}\label{rhovsr-h}
  \end{center}
\end{figure}

A very important piece of information is the way in which the transition is made. If second order, the chemical potential should be flat at the transition, identified as the point where the chemical potential changes sign. Let us consider the isothermal functions of chemical potential versus density. In Fig. \ref{nvsmu-FD} we show an actual curve $\mu$ vs $n$ for  a temperature $T = 1.0$ for an ideal Fermi gas, while in Fig. \ref{muvsn-h} we show a {\it hypothetically expected} isotherm for an interacting gas, be it Fermi or Bose. The main difference between the latter and the former is that the slope of $\mu$ vs $n$ at the transition $\mu = 0$ is zero, as dictated by the fact that it should be a second order phase transition. The tail for $\mu < 0$ should become that of an ideal {\it classical} gas, while the corresponding one for $\mu > 0$ should follow that of Bogolubov or BCS theories, as the temperature nears zero.  The difference of those curves being quantitative but certainly not qualitative. Experiments with ultracold gases have the peculiarity that are performed in confining potentials, typically harmonic, that give raise to inhomogenous gases. It is well established that the so-called {\it local density approximation} (LDA) yields accurately the density profile measured in these experiments\cite{Dalfovo,VRR1,Horikoshi,Ho,Salomon,VRR2}. This approximation indicates that the density profile, for an isotropic harmonic potential, may be calculated as
\begin{equation}
\rho(r) = n(\mu - \frac{1}{2}m\omega^2 r^2,T) .\label{profile}
\end{equation}
That is, the equation of state $n = n(\mu,T)$ for the homogenous system is the inhomogenous density profile rescaled by the confining potential. Fig. \ref{rhovsr-h} is the profile for an isotherm of the hypothetical interacting system of Fig. \ref{muvsn-h}, for a temperature and chemical potential of the confined system below condensation. A very important prediction of the present study is thus found: the point where the chemical potential is zero in the {\it homogenous} system translates to the point where the profile changes behavior. Call this spatial point $r_c$, then the chemical potential of the {confined inhomogenous} gas is $\mu = \frac{1}{2} m \omega^2 r_c^{2}$.  That is, density profiles are enough to construct not only the phase diagram of the homogeneous systems but also of the inhomogenous ones. The latter follow from the fact that the number of particles in the confined gas is
\begin{equation}
N = \int \rho(r) d^3 r .
\end{equation}
This yields $N = N(T,\mu, \omega^{-3})$ which is the equation of state of the {\it confined} gas as a function of $T$, $\mu$ and the ``generalized volume" $\omega^{-3}$\cite{VRR1,VRR2}; see also Ref. \cite{BRAZIL} where an effort has been made to experimentally construct the phase diagram of a trapped $^{87}$Rb gas. Although finite size effects may obscure the behavior of the density profiles at the region where they typically change curvature, systematic measurements with different number of particles should help to elucidate this relevant point. 

We recall here that building phase diagrams from density profiles has already been performed\cite{Horikoshi,Ho,Salomon}, however, the identification of zero chemical potential as the transition points has not been used.

\section{Final Remarks}

In his celebrated book\cite{LondonBook} on superfluidity, Fritz London makes a description of the nature of the superfluid phase that essentially has survived up to these days, and that it is  also the main point of the present article although with a twist. London was impressed by the fact that the superfluid fraction does not contribute to the entropy of the fluid. He considers the fraction to be a single macroscopic quantum state and he even says that superfluid flow allows to ``tap ground state" from the fluid\cite{LondonBook}. The further but inaccurate London's assumption is that the superfluid state may be considered to be similar to an ideal Bose-Einstein condensate. It is well recorded that Landau\cite{LandauPR41} argued that it could not be so since an ideal gas cannot be superfluid. Few years later, Bogolubov\cite{Bogolubov} provided the first satisfactory model of superfluidity at $T = 0$, which needed the presence of intermolecular interactions in an essential way. Nevertheless, since then, many of theoretical efforts and explanations have been prone to describe superfluidity as a kind of Bose-Einstein condensation in a single particle state, although certainly including interactions. It may seem, as London himself believed, that Bose statistics are essential. However, we also well know that this is not necessarily true. BCS-like theories have been successfully applied\cite{LeggettPRL72} to explain superfluidity in $^3$He, a fermion. Thus, the existence of quantum fluids phases that may exhibit superfluidity is not truly a matter of statistics, it is a phenomenon due solely to the quantum nature of the fluids and to the fact that the samples are macroscopic. Of course, the details may depend on the particular fluid, and thus certain properties may depend on being bosons or fermions. And it is in this regard that the main thesis of this article rests: the phenomenon of condensation is the sudden occupation of certain number of atoms of {\it their} many-body ground state, hence  not contributing to the entropy, and reaching the full ground state of the whole sample at zero temperature. Clearly, this scenario includes ideal Bose-Einstein condensation, but it does appear that ideal Fermi-Dirac condensation is closer to the interacting case, be it fermions or bosons. One of the suggestions of this article is that such an observation may be useful in the search of a complete and satisfactory theory of the thermodynamics of quantum fluids.

\begin{acknowledgments}

We acknowledge support from grant PAPIIT-UNAM IN116110. I thank Dulce Aguilar for technical assistance in the preparation of the figures of this article.

\end{acknowledgments}

\end{document}